# Dynamically Multivalued Self-Organisation and Probabilistic Structure Formation Processes

A.P. Kirilyuk[1]

Institute of Metal Physics, pr. Vernadskogo 36, Kiev-142, Ukraine 03142

**Keywords:** Complexity, chaos, information, entropy, fractal, adaptability, nanotechnology

**Abstract.** The unreduced, universally nonperturbative analysis of arbitrary many-body interaction process reveals the irreducible, purely dynamic source of randomness. It leads to the universal definition of real system complexity, where the internally chaotic self-organisation emerges as a characteristic case of complex interaction dynamics. One obtains the causally complete description of the world structure emergence, from elementary particles to consciousness, including practically important and fundamentally substantiated propositions for self-formation research and applications.

**Introduction**

Any structure formation process originates in the underlying interactions. A generic interaction is described mathematically by a *nonintegrable* dynamic equation, which constitutes the *unreduced* formulation of the many-body problem. Not only it cannot be solved within the conventional theory framework, but the latter does not provide, in general, any sound ideas about the true meaning of "nonintegrability". Various approximations, always reduced to a version of *perturbation theory*, are then used to transform the problem into an integrable *model* of the real system that does possess a closed, or "exact", solution, but lacks some essential properties of structure formation, among which the *autonomous* system *creativity*, i.e. explicit structure emergence without its artificial insertion in another form, is of particular interest for understanding of self-formation processes, ubiquitous in living nature and now also intensively studied in artificially designed, often small-scale systems.

In this report we describe a recently elaborated, universally applicable method of solution of the unreduced many-body problem and show that the obtained nonperturbative, truly *general* problem solution possesses the quality of autonomous creativity, inevitably absent in the conventional, reduced models and perturbative solutions. This essential quality of real structure formation emerges together with the *universally* valid concept of *dynamic complexity* and *chaoticity* based on the key property of *dynamic multivaluedness* and *entanglement* of the unreduced problem solution missing in its reduced models (including the conventional versions of "complexity", "chaoticity", and "self-organisation") [1-5]. We then describe the main properties of the unreduced dynamic complexity and the ensuing applications to various natural and artificial structure formation processes.

**Unreduced Analysis of Arbitrary Many-Body Interaction Problem**

We start with the most general formulation of a problem of interaction within a many-component system [2], in the form of *existence equation* that generalises various particular equations:

$$\left\{ \sum_{k=0}^{N} \left[ h_k(q_k) + \sum_{l>k}^{N} V_{kl}(q_k,q_l) \right] \right\} \Psi(Q) = E\Psi(Q) , \qquad (1)$$

where $h_k(q_k)$ is the "generalised Hamiltonian" for the $k$-th system component in the absence of interaction, $q_k$ denotes the degree(s) of freedom of the $k$-th component, $V_{kl}(q_k,q_l)$ is the "interaction potential" between the $k$-th and $l$-th components, $\Psi(Q)$ is the system "state function" characterising completely the compound system state and depending on the degrees of freedom of all interacting components, $Q \equiv \{q_0, q_1, ..., q_N\}$, $E$ is the "eigenvalue" of the generalised Hamiltonian in the state $\Psi(Q)$, and summations are performed over all system components. As shown in the universal science of complexity [1,2], any correct equation is reduced to a Hamiltonian form of Eq. 1, which actually expresses only the fact of interaction (initial system configuration). It can often be useful to separate one of the participating degrees of freedom, say $q_0 \equiv \xi$, from other variables $q_k$ ($k = 1,...,N$):

$$\left\{ h_0(\xi) + \sum_{k=1}^{N} [h_k(q_k) + V_{0k}(\xi, q_k)] + \sum_{l>k}^{N} V_{kl}(q_k, q_l) \right\} \Psi(\xi, Q) = E\Psi(\xi, Q) , \qquad (2)$$

where $Q \equiv \{q_1,...,q_N\}$ and $k, l \geq 1$. The separated variable $\xi$ can describe "common" (extended) degree(s) of freedom, providing physical "coordinates" for other, localised elements.

---
[1] Address for correspondence: Post Box 115, Kiev-30, Ukraine 01030. E-mail address: kiril@metfiz.freenet.kiev.ua .

It is convenient to express the problem in terms of the known solutions for the free components:

$$h_k(q_k)\varphi_{kn_k}(q_k) = \varepsilon_{n_k}\varphi_{kn_k}(q_k) , \qquad (3)$$

where $\{\varphi_{kn_k}(q_k)\}$ are the eigenfunctions and $\{\varepsilon_{n_k}\}$ eigenvalues of the $k$-th component Hamiltonian $h_k(q_k)$, forming the complete set of orthonormal functions. Expanding the total system state-function $\Psi(q_0, q_1,...,q_N)$ over complete sets of eigenfunctions $\{\varphi_{kn_k}(q_k)\}$ for the "functional" degrees of freedom $(q_1,...,q_N) \equiv Q$, which describe the "internal" dynamics of system elements, we are left with functions depending only on the selected "structural" degrees of freedom $q_0 \equiv \xi$:

$$\Psi(q_0, q_1,...,q_N) \equiv \Psi(\xi, Q) = \sum_{n \equiv (n_1, n_2,...,n_N)} \psi_n(q_0)\varphi_{1n_1}(q_1)\varphi_{2n_2}(q_2)...\varphi_{Nn_N}(q_N) \equiv \sum_n \psi_n(\xi)\Phi_n(Q) , \qquad (4)$$

where the summation is performed over all possible combinations of eigenstates $n \equiv (n_1, n_2,...,n_N)$ and we have designated $\Phi_n(Q) \equiv \varphi_{1n_1}(q_1)\varphi_{2n_2}(q_2)...\varphi_{Nn_N}(q_N)$ for brevity. The system of equations for $\{\psi_n(\xi)\}$ is obtained from Eq. 2 after substitution of expansion of Eq. 4, multiplication by $\Phi_n^*(Q)$ and integration over all variables $Q$ (using the eigenfunction orthonormality):

$$H_0(\xi)\psi_0(\xi) + \sum_n V_{0n}(\xi)\psi_n(\xi) = \eta\psi_0(\xi) , \qquad (5a)$$

$$H_n(\xi)\psi_n(\xi) + \sum_{n' \neq n} V_{nn'}(\xi)\psi_{n'}(\xi) = \eta_n\psi_n(\xi) - V_{n0}(\xi)\psi_0(\xi) , \qquad (5b)$$

where $n, n' \neq 0$ (also everywhere below), $\eta \equiv \eta_0 \equiv E - \varepsilon_0$,

$$\eta_n \equiv E - \varepsilon_n , \quad \varepsilon_n \equiv \sum_k \varepsilon_{n_k} , \quad H_n(\xi) = h_0(\xi) + V_{nn}(\xi), \quad V_{nn'}(\xi) = \sum_k \left[ V_{k0}^{nn'}(\xi) + \sum_{l>k} V_{kl}^{nn'} \right] , \qquad (6)$$

$$V_{k0}^{nn'}(\xi) = \int_{\Omega_Q} dQ \Phi_n^*(Q) V_{k0}(q_k, \xi) \Phi_{n'}(Q), \quad V_{kl}^{nn'}(\xi) = \int_{\Omega_Q} dQ \Phi_n^*(Q) V_{kl}(q_k, q_l) \Phi_{n'}(Q) , \qquad (7)$$

and we have separated the equation for $\psi_0(\xi)$ describing the usually measured generalised "ground state" of the system elements, i.e. the state with minimum energy and complexity.

In each equation for $\psi_n(\xi)$, Eqs. 5b, we express $\psi_n(\xi)$ through $\psi_0(\xi)$ using the usual Green function for the "homogeneous" equation part [1-6]. Substituting the result into Eq. 5a, we get the *effective* existence equation for $\psi_0(\xi)$:

$$h_0(\xi)\psi_0(\xi) + V_{\text{eff}}(\xi;\eta)\psi_0(\xi) = \eta\psi_0(\xi) , \qquad (8)$$

where the operator of the *effective (interaction) potential (EP)*, $V_{\text{eff}}(\xi;\eta)$, is given by

$$V_{\text{eff}}(\xi;\eta) = V_{00}(\xi) + \hat{V}(\xi;\eta) , \quad \hat{V}(\xi;\eta)\psi_0(\xi) = \int_{\Omega_\xi} d\xi' V(\xi,\xi';\eta)\psi_0(\xi') , \qquad (9a)$$

$$V(\xi,\xi';\eta) = \sum_{n,i} \frac{V_{0n}(\xi)\psi_{ni}^0(\xi)V_{n0}(\xi')\psi_{ni}^{0*}(\xi')}{\eta - \eta_{ni}^0 - \varepsilon_{n0}} , \qquad \varepsilon_{n0} \equiv \varepsilon_n - \varepsilon_0 , \qquad (9b)$$

and $\{\psi_{ni}^0(\xi)\}$, $\{\eta_{ni}^0\}$ are the complete sets of eigenfunctions and eigenvalues, respectively, for the truncated system of equations obtained as "homogeneous" parts of Eqs. 5b:

$$H_n(\xi)\psi_n(\xi) + \sum_{n'\neq n} V_{nn'}(\xi)\psi_{n'}(\xi) = \eta_n\psi_n(\xi) . \qquad (10)$$

The eigenfunctions, $\{\psi_{0i}(\xi)\}$, and eigenvalues, $\{\eta_i\}$, of Eq. 8 are then used to obtain other state-function components:

$$\psi_{ni}(\xi) = \hat{g}_{ni}(\xi)\psi_{0i}(\xi) \equiv \int_{\Omega_\xi} d\xi' g_{ni}(\xi,\xi')\psi_{0i}(\xi'), \quad g_{ni}(\xi,\xi') = V_{n0}(\xi')\sum_{i'} \frac{\psi_{ni'}^0(\xi)\psi_{ni'}^{0*}(\xi')}{\eta_i - \eta_{ni'}^0 - \varepsilon_{n0}} , \qquad (11)$$

after which the total system state-function $\Psi(q_0,q_1,...,q_N) = \Psi(\xi, Q)$, Eq. 4, is obtained as

$$\Psi(\xi,Q) = \sum_i c_i \left[ \Phi_0(Q) + \sum_n \Phi_n(Q)\hat{g}_{ni}(\xi) \right] \psi_{0i}(\xi) , \qquad (12)$$

where the coefficients $c_i$ should be found by state-function matching at the boundary/moment where the effective interaction vanishes. The observed (generalised) density, $\rho(\xi,Q)$, is obtained as the state-function squared modulus, $\rho(\xi,Q) = |\Psi(\xi,Q)|^2$ (for "quantum" and other "wave-like" levels of complexity), or as the state-function itself, $\rho(\xi,Q) = \Psi(\xi,Q)$ (for "particle-like" levels) [1].

Although the "effective" problem expression, Eqs. 8-12, is formally equivalent to its initial formulation, Eqs. 1,2,5, it reveals the qualitatively new phenomenon of *dynamic multivaluedness (or redundance)*, which appears as a redundant number of locally complete, and therefore *mutually incompatible*, solutions of the effective existence equation, Eqs. 8-9, called system *realisations* (each of them represents a completely defined system configuration). As all realisations are *equally* possible and real, the system is forced, by the driving interaction itself, to permanently change its realisations in a *causally random* order. The multiplicity of realisations follows from the elementary calculation of the eigen-solution number determined by the highest power of the characteristic equation for Eqs. 8-9 [1-6]. If $N_\xi$ and $N_q$ are the numbers of terms in the sums over $i$ and $n$ in Eq. 9b, equal, respectively, to the numbers of system components ($N$) and their internal states, then the total eigenvalue number is $N_{\max} = N_\xi(N_\xi N_q + 1) = (N_\xi)^2 N_q + N_\xi$, giving the $N_\xi$-fold redundancy of the usual "complete" set of $N_\xi N_q$ eigen-solutions of Eqs. 5 plus an additional, "incomplete" set of $N_\xi$ eigen-solutions. Therefore the total number of "regular", redundant system realisations is $N_\Re = N_\xi = N$, whereas the mentioned additional set of solutions forms the specific, "intermediate" realisation that plays the role of transitional state during chaotic system jumps between the "regular" realisations and provides thus the universal, *causally complete* extension of the quantum-mechanical *wavefunction* and classical *distribution function* [1-3]. Note that the "effective" problem formulation is well known under the name of "optical, or effective, potential method" (see e. g. [7]), but it is invariably used in its perturbative, reduced version that "kills" dynamic multivaluedness and corresponds to unrealistic, *dynamically single-valued*, or *unitary*, system "model" devoid of any intrinsic creativity (the single remaining realisation corresponds usually to the intermediate realisation, i. e. generalised distribution function, representing a statistically averaged, *effectively one-dimensional, or even zero-dimensional, projection* of unreduced, multivalued system dynamics).

The derived dynamic multivaluedness and randomness of the unreduced, general problem solution means that the measured system density, $\rho(\xi,Q)$, should be presented as the special, *causally, or dynamically, probabilistic* sum of the corresponding densities for individual realisations, $\{\rho_r(\xi,Q)\}$, obtained by solution of the effective existence equation, Eqs. 8-9:

$$\rho(\xi,Q) = \sum_{r=1}^{N_\Re} {}^\oplus \rho_r(\xi,Q), \tag{13}$$

where the summation is performed over all the *actually observed* system realisations, numbered by $r$, and the sign $\oplus$ serves to designate the special, dynamically probabilistic meaning of the sum derived above and consisting in *permanent change* of regular system realisations in *dynamically random (chaotic) order* by passing each time through the intermediate realisation. The dynamically obtained, *a priori probability* of the $r$-th actually observed realisation, $\alpha_r$, is determined, in general, by the number, $N_r$, of elementary, experimentally unresolved realisations it contains:

$$\alpha_r(N_r) = \frac{N_r}{N_\Re} \quad \left(N_r = 1,...,N_\Re; \sum_r N_r = N_\Re\right), \qquad \sum_r \alpha_r = 1. \tag{14}$$

According to "generalised Born's rule" that follows from the dynamical matching conditions naturally realised in the intermediate realisation (wavefunction), the dynamic probability values are practically determined by the values of the generalised wavefunction (or distribution function) obeying the causally extended, *universal Schrödinger equation* [1,2].

The property of *dynamic entanglement* of the interacting system components is closely related to dynamic multivaluedness and appears in the unreduced solution, Eq. 12, as products of functions of $Q$ and $\xi$ "weighted" by the dynamically determined factors which account for the physically transparent *nonseparability* of the entangled configuration of each system realisation (however, the system components *transiently* disentangle during its chaotic jumps between realisations [1-4]). This nonseparability of the dynamically multivalued entanglement only increases due to the hierarchically reproduced, *dynamically fractal* structure of the developing interaction process, where the solutions of the truncated system of equations, Eqs. 10, show the same properties of dynamic multivaluedness and entanglement obtained by the same EP method and reproduced at all naturally emerging levels of structure formation [1-4]. This *dynamically probabilistic fractal* extends considerably the conventional fractality concept and contributes essentially to unreduced interaction *creativity*, since finer parts of its distributed structure provide the causally obtained "interaction potential" for the next level of structure formation, while the *interaction-driven* randomness accounts for the autonomous, *dynamic adaptability* of the emerging "living" structures.

**Universal Dynamic Complexity, Its Particular Cases, Symmetry and Applications**

After the unreduced interaction analysis has *explicitly* given the complete set of realisations, the *dynamic complexity* itself, $C$, is *universally* defined as any growing function of (observed) realisation number, $C = C(N_\Re)$, $dC/dN_\Re > 0$, or the rate of their change, equal to zero for the (unrealistic) case of only one realisation, $C(1) = 0$. It is just that, unrealistically simplified "model" (zero-dimensional, point-like projection) of reality which is *exclusively* considered in the conventional, unitary science (including its concepts of "complexity", "chaoticity", "self-organisation", etc.), which explains all its persisting "difficulties", "mysteries", and "insoluble problems", easily finding their natural, dynamically multivalued solution by the unreduced complexity concept [1-6] that emerges as the direct *extension* of the unitary knowledge model to the dynamically multivalued reality.

Two limiting, characteristic cases of the unreduced interaction dynamics are then obtained, together with the inhomogeneous transition between them [1,2]. If the parameters (universally expressed by frequencies) of the interacting entities and interaction itself are close to each other, then the regime of *uniform, or global, chaos* emerges, characterised by a homogeneous distribution of realisation probabilities (i. e. $N_r \approx 1$ and $\alpha_r \approx 1/N_\Re$ for all $r$ in Eq. 14) and the corresponding, "very irregular" system behaviour. If there is essential parameter difference within the system, then the regime of *dynamically multivalued self-organisation, or self-organised criticality (SOC)* sets in, where the realisation probability distribution is essentially inhomogeneous, and the (few) emerging low-frequency, externally "regular" structures contain, or "enslave", (many) high-frequency, chaotically changing constituents. By contrast to the conventional self-organisation and similar concepts, the unreduced, real-world SOC *always* has the chaotic (multivalued) and fractal (hierarchic) internal structure and origin. It provides also the *intrinsically unified* extension of the whole diversity of similar concepts, remaining irreducibly separated and incomplete within the unitary "science of complexity", such as self-organisation (synergetics), SOC, "control of chaos", "synchronization", "attractors", and "mode locking". Moreover, the permanently localised, trajectorial behaviour from the "classical" ("Newtonian") science is *causally obtained* as a particular case of the *internally chaotic* SOC, while the laws of classical mechanics (and their relativistic extension) can now be *consistently derived* (and extended) in terms of the *underlying unreduced complexity* [1,2], instead of their imposition by formal, flawed postulates in the unitary theory.

The *universal criterion of global (uniform) chaos* onset is obtained in the form [1,2]

$$\kappa \equiv \frac{\Delta \eta_i}{\Delta \eta_n} = \frac{\omega_\xi}{\omega_q} \cong 1 \ , \tag{15}$$

where $\kappa$ is the introduced *chaoticity* parameter, $\Delta \eta_i$, $\omega_\xi$ and $\Delta \eta_n \sim \Delta \varepsilon$, $\omega_q$ are energy-level separations and frequencies for the inter-component and intra-component motions, respectively. At $\kappa \ll 1$ one has the multivalued SOC regime (showing external "regularity"), which becomes the less and less regular as $\kappa$ grows from 0 to 1, until at $\kappa \approx 1$ the global chaos sets in, followed by another SOC regime of a "reversed" system configuration at $\kappa \gg 1$. Note that the simple and universally applicable chaos (and SOC) criterion of Eq. 15 determines the qualitative kind of *any* system behaviour and reveals the true meaning and result of the "well-known" phenomenon of resonance, as opposed to multiple separated, technically complicated, abstract-model dependent, and inconsistent chaos criteria and "signatures" of the unitary "science of complexity" that cannot avoid confusion while comparing its results with real system behaviour (see e. g. Refs. [8,9]).

The unified theory of unreduced dynamic complexity culminates in the rigorously substantiated, *universal law of symmetry, or conservation, of complexity* [1,2], which includes the causally extended, complex-dynamical versions of *all* the (correct) laws and "principles" of the unitary science and states that the *total*, unreduced system complexity is always *preserved*, but *only* by way of its *unceasing transformation* (development, unfolding) from the latent, "potential" form of *dynamic information* (universally expressed by the generalised mechanical *action*) to the explicit, unfolded form of *dynamic entropy*. It is important that now *any* process, including emergence of the most externally "regular" structure, is definitely associated with the dynamic entropy *growth*, which resolves bundles of stagnating problems of the conventional science (including problems around time and life). Contrary to "regular", but "broken" symmetries of the unitary science, the symmetry of complexity is *always exact* (never "broken"), but reveals symmetric relations between and within always somewhat *different, irregular* structures and dynamic regimes.

The experience of the already realised applications of the universal science of complexity, from fundamental physics to the theory of consciousness and humanitarian fields [1-6], confirm the rigorously substantiated conclusions that have the direct, practical importance for the emerging science of autonomously developing, "self-forming" systems of various origins. First of all, the proposed theory of *unreduced, reality-based* dynamic complexity in terms of *dynamically multivalued* entanglement of interacting system components appears to be indispensable for any useful understanding of self-formation processes and cannot be replaced in this case by any unitary (dynamically single-valued) imitation. Second, one cannot, and should not, avoid the true, dynamic randomness in the system behaviour, based on the irreducible dynamic multivaluedness of the driving interaction process and playing the major, *creative* role in the system structure formation (see Ref. [2] for the detailed application to popular cases of quantum computation and nanotechnology). And finally, the unified, universally confirmed law of conservation (symmetry) of complexity shows that every "miracle" of new structure emergence is always obtained at a well-defined, corresponding cost, and one can get only those "self"-organised structures that have been properly conceived in terms of the driving interaction parameters (in accord with the above rigorous interaction analysis).